\documentclass[journal=jacsat,manuscript=article]{achemso}
\usepackage[version=3]{mhchem} 
\usepackage{color}
\usepackage{url}
\newcommand{\mc}[3]{\multicolumn{#1}{#2}{#3}}
\newcommand{\onlinecite}[1]{\hspace{-1 ex} \nocite{#1}\citenum{#1}}
\title{How Cations Change Peptide Structure}

\author{Carsten Baldauf}
\email{baldauf@fhi-berlin.mpg.de}
\affiliation[Fritz-Haber-Institut der MPG]{Fritz-Haber-Institut der Max-Planck-Gesellschaft, Faradayweg 4-6, D-14195 Berlin-Dahlem, Germany}
\author{Kevin Pagel}
\email{pagel@fhi-berlin.mpg.de}
\affiliation{Fritz-Haber-Institut der Max-Planck-Gesellschaft, Faradayweg 4-6, D-14195 Berlin-Dahlem, Germany}
\author{Stephan Warnke}
\affiliation{Fritz-Haber-Institut der Max-Planck-Gesellschaft, Faradayweg 4-6, D-14195 Berlin-Dahlem, Germany}
\author{Gert von Helden}
\affiliation{Fritz-Haber-Institut der Max-Planck-Gesellschaft, Faradayweg 4-6, D-14195 Berlin-Dahlem, Germany}
\author{Beate Koksch}
\affiliation[Freie Universit\"at Berlin]{Institut f\"ur Chemie und Biochemie - Organische Chemie, Freie Universit\"at Berlin, Takustr. 3, D-14195 Berlin-Dahlem, Germany}
\author{Volker Blum}
\email{blum@fhi-berlin.mpg.de}
\affiliation{Fritz-Haber-Institut der Max-Planck-Gesellschaft, Faradayweg 4-6, D-14195 Berlin-Dahlem, Germany}
\author{Matthias Scheffler}
\affiliation{Fritz-Haber-Institut der Max-Planck-Gesellschaft, Faradayweg 4-6, D-14195 Berlin-Dahlem, Germany}

\date{\today}
\begin{document}
\begin{abstract}
Specific interactions between cations and proteins have a strong impact on peptide and protein structure. 
We here shed light on the nature of the underlying interactions, especially regarding the  effects on the polyamide backbone structure. 
To do so, we compare the conformational ensembles of model peptides in isolation and in the presence of either \ce{Li+} or \ce{Na+} cations by state-of-the-art density-functional theory (including van der Waals effects) and gas-phase infrared spectroscopy.
These monovalent cations have a drastic effect on the local backbone conformation of turn-forming peptides, by disruption of the H bonding networks and the resulting severe distortion of the backbone conformations. 
In fact, \ce{Li+} and \ce{Na+} can even have different conformational effects on the same peptide.
We also assess the predictive power of current approximate density functionals for peptide-cation systems and compare to results from established protein force fields as well as to high-level quantum chemistry (CCSD(T)).
\end{abstract}

\maketitle

\section{Introduction}
As early as 1912, Paul Pfeiffer systematically studied the crystallization of short Ala and Gly containing peptides from aqueous solution in the presence of alkali salts\cite{pfeiffer1912} and postulated that \ce{Li+} exhibits a higher affinity (``Additionsf\"ahigkeit'') to peptides than \ce{Na+} and \ce{K+}.\cite{pfeiffer1924} 
Indeed, calorimetric studies revealed high interaction enthalpies of a series of peptides with \ce{Li+},\cite{seebach1994calorimetric} in the range of solvation enthalpies of peptides. 
These strong interactions are in practice used to increase the share of \textit{cis} prolyl peptide bonds from 10\% to 70\% upon addition of \ce{Li+}-salts in biochemical assays of the activity of peptidyl prolyl \textit{cis}-\textit{trans} isomerases.\cite{kofron1991determination}
NMR studies of the cyclic peptide Cyclosporin A (CysA) in organic solvents revealed that \ce{Li+} inhibits the formation of H bonds and induces 'unusual' backbone conformations.\cite{kessler1990complexation,kock1992novel} 
100 years after Pfeiffer's work, Garand \textit{et al.}\cite{Garand10022012} studied the non-covalent interactions of a non-natural peptide-based catalyst by means of gas-phase infrared (IR) spectroscopy.
While the polyamide backbone of the molecule forms intramolecular H bonds if protonated, sodiation apparently results in a complete absence of H bonds due to the involvement of the carbonyl groups in interactions with the \ce{Na+} cation.
Both CysA \cite{kessler1990complexation,kock1992novel} and the peptide-based catalyst \cite{Garand10022012} form narrow turn-like backbone loops, well suited to accommodate a cation.  
Such turns are normally at the outside of globular proteins, exposed to the surrounding medium. 
We here investigate the atomistic and electronic basis of 
cation peptide interactions in turn-forming peptides.
The focus is on proline-containing peptides, where pronounced conformational effects of such interactions can be expected due to the possible \textit{cis} and \textit{trans} states of the prolyl-peptide bond.\cite{ernst1995modern} 
Our study is based on accurate conformational predictions by first-principles (density-functional theory) in a synergistic combination with gas-phase IR spectroscopy to validate the results.

Structure formation and dynamics in proteins can be primarily attributed to the rotation of the N--C$\alpha$ and C$\alpha$--C single bonds, represented by the backbone torsion angles $\phi$ and $\psi$, respectively (Figure~1A).
This conformational $\phi$/$\psi$ space is well described by a Ramachandran diagram \cite{ramachandran1963stereochemistry} like the statistical evaluation of high-resolution X-ray data \cite{lovell2003structurevalidation} shown in Figure~1B.
The blue shaded areas are referred to as allowed conformational regions and can be associated with characteristic secondary structure types (Figure~1B).

The double bond character of the peptide bond hinders free rotation and allows for two distinct conformations.
In general, the \textit{trans} conformation with an apparently high barrier to \textit{cis} is almost exclusively observed.\cite{fischer2000chemical,Dugave2003}
A significant fraction of \textit{cis} conformation is only observed for the prolyl-peptide bond.\cite{weiss_peptide_1998}
In proline, \textit{cis} and \textit{trans} (Figure~1C) are close in energy since the C$\beta$ of the preceding building block encounters a carbon atom of proline (C$\alpha$ or C$\delta$) in both states.
A \textit{cis} peptide bond, usually preceding a proline building block, is a feature of so-called type $\beta VI$  turns (Figure~1D).\cite{richardson_anatomy_1981,moehle_structural_1997}
This notation dates back to Venkatachalam, according to which $\beta$-turns share the feature of an H-bond between residues $i+3 \rightarrow i$ and are further classified by the backbone torsion angles $\phi$ and $\psi$ of the residues $i+1$ and $i+2$.\cite{venkatachalam_stereochemical_1968} 
The $\beta$ turns of the protein backbone allow for a 180$^\circ$ reversal of the direction of structure propagation within four consecutive residues of a polypeptide chain.
Similarly, Hutchinson and Thornton classify $\beta$-turns according to ranges for the backbone torsion angles $\phi$ and $\psi$ into eight well-defined classes ($I$, $I'$, $II$, $II'$, $VIa1$, $VIa2$, $VIb$, and $VIII$) and a miscellaneous type $IV$.\cite{sibanda_b-hairpin_1985, hutchinson_revised_1994}
Very prominent are the common (type $I$) and Glycine (type $II$) turn and their inverse counterparts $I'$ and $II'$.
The special $\beta$-turn types $VIa$ and $VIb$ have a \textit{cis} peptide bond between central residues $i+1$ and $i+2$; they frequently feature proline in position $i+2$.\cite{richardson_anatomy_1981}
Idealized backbone torsion angles for these turn types can be found, e.g., in reference~\onlinecite{moehle_structural_1997}.

\begin{figure}
 \includegraphics[width=0.44\textwidth]{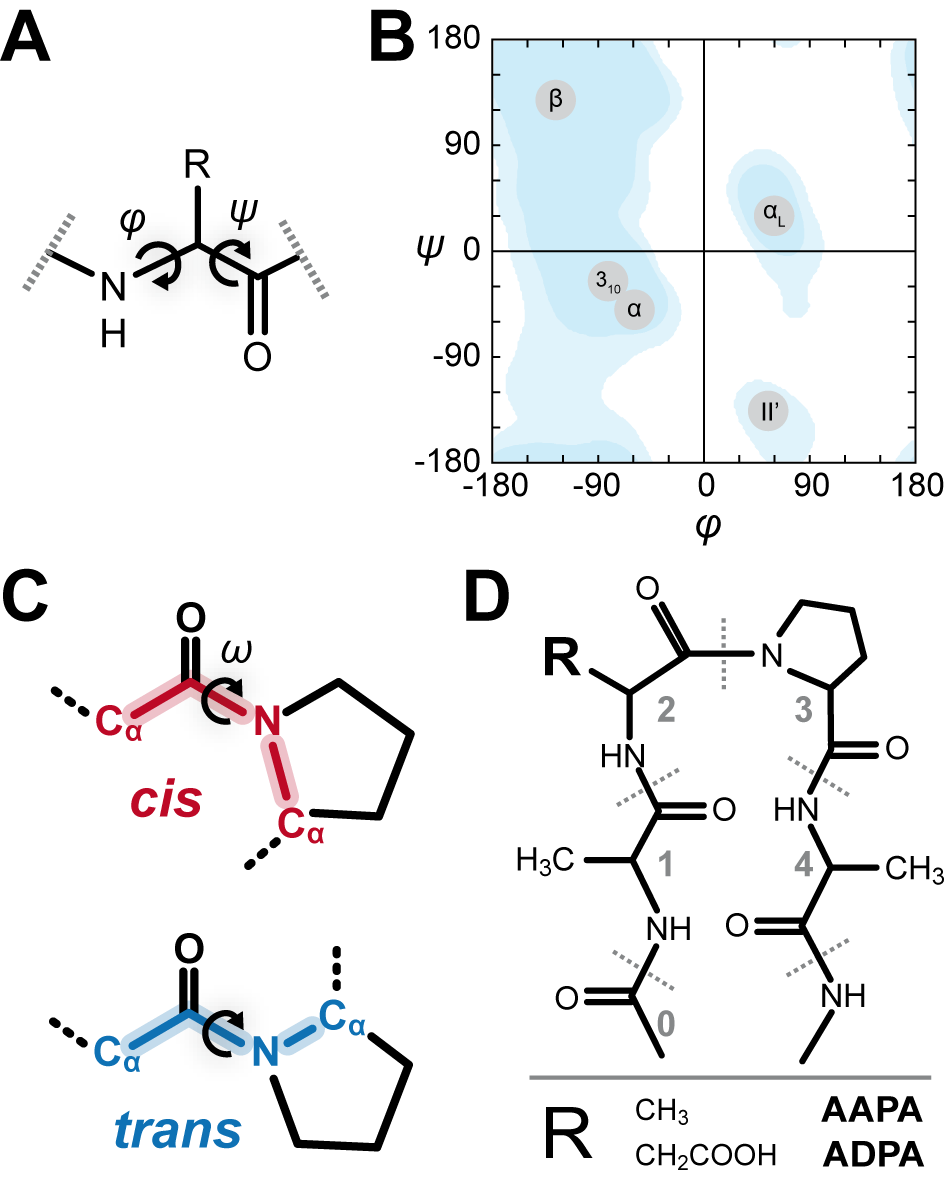}
 \caption{The backbone torsion angles $\phi$ and $\psi$ (\textbf{A}) of the residues of a polypeptide chain can be illustrated by a Ramachandran plot (\textbf{B}), based on data from reference \cite{lovell2003structurevalidation}. Labels highlight characteristic secondary structure types: The $\beta$ region in the 2nd quadrant, the $3_{10}$ and $\alpha$ helical region in the 3rd quadrant as well as left-handed $\alpha$ and the $\beta$II' region in the 1st and 4th quadrant, respectively. The \textit{cis} and \textit{trans} state of the prolyl-peptide bond (\textbf{C}) has a drastic effect on polypeptide structure, which we study with the model peptides AAPA and ADPA (\textbf{D}), here shown in a schematic $\beta VI$-turn conformation.
}
 \label{fig:Intro_Formula}
\end{figure}

In this study, we make use of the characteristic of proline-containing peptides to allow for the formation of \textit{cis} and \textit{trans} peptide bonds as a potential strong ``conformational signal`` triggered by the peptide cation interaction.
Indeed, Seebach and co-workers reported ion-induced conformational effects on peptide structure to be especially pronounced in the proximity of proline.\cite{ernst1995modern}
Kunz \textit{et al.} investigated a systematic series of proline-containing peptides with NMR and find peptides containing an Asp-Pro sequence to exhibit a \textit{cis}/\textit{trans} ratio that is in opposition to the results of all other sequences studied.\cite{kunz2012lithium}
Therefore, we here investigate the sequence AXPA (Figure~1D), where P is the single letter code for proline and X is either alanine (A) or aspartate (D), allowing us to differentiate between pure cation backbone interaction and the additional effect of cation sidechain interactions on the peptide backbone conformation.
The peptide design in the present work avoids structure-perturbing labels and protects the termini with acetyl and aminomethyl groups (Figure~1D) to embed the sequence in a 'protein-like' chain structure, avoiding endgroup effects, i.e., zwitterion formation.

\section{Results and discussion}
We combine an exhaustive conformational search from first principles with theoretical and experimental gas-phase IR spectroscopy.
Such investigations of isolated peptides in the gas phase offer an unbiased view of structure formation trends intrinsic to the molecule, a successful strategy for charged and uncharged amino acids and peptides.\cite{aboriziq2006gasphase,bakker2003fingerprint,compagnon2006peptides,kamariotis2006infrared,cimas2009vibrational,james2009intramol,james2010laserspec,rossi2010secondary,plowright2011compact,chutia2012adsorption,rossi2013vibrations}
By the stepwise addition of perturbing contributions, in this case cations, we aim to reconstruct the main contributions to protein secondary structure formation in a bottom-up approach.
The success of such an approach is critically linked to the quality of the description of the potential-energy surface of the system under investigation.
We employ density-functional theory (DFT) in the generalized-gradient approximation with the Perdew-Burke-Ernzerhof (PBE) functional.\cite{perdew1996generalized}
Van der Waals dispersion interactions are included through a pairwise $C_6 R^{-6}$ term for which $C_6$ coefficients are derived from the self-consistent electron density, referred to as PBE+vdW.\cite{tkatchenko2009accurate} 
Our use of rather accurate, but computational efficient approximate DFT is justified by the high-level benchmarks we present in the methods section of this article. 

\subsection{Conformational analysis}

The theoretical conformational analysis of the short peptide AAPA (Figure~1D) is already challenging.
Hypothetically, discretizing the backbone torsion angles with a 30 degree grid and assuming two possible states (\textit{cis} and \textit{trans}) for the peptide bonds would formally result in roughly 35 million conformations to evaluate. 
In order to tackle this massive conformational space, we resort to a basin hopping-like exhaustive search of the potential-energy surfaces (PES) of conventional protein force fields (OPLS-AA \cite{jorgensen2004free} or AMBER99 \cite{wang2000well}). 
We employ the TINKER~5 scan routine \cite{pappu1998analysis} in an in-house parallelized version.
In order to achieve a reliable and parameter-free description, we then follow up with a large set (700 to 1800 per peptide-cation system) of PBE+vdW post-relaxation calculations as a second computational step.

Figure~2 shows our results for AAPA in isolation. The lowest-energy structure of the PES, a $\beta VI$ turn with a \textit{cis} prolyl peptide bond, also has the lowest free energy in the harmonic approximation.
Two alternative $\beta VI$ turns are 4.5 and 8.3\,kJ/mol higher in $\Delta F_{300K}$.
The most stable conformer with a \textit{trans} peptide bond is a $\beta II$' turn with $\Delta E$=2.8\,kJ/mol.
Harmonic free energy contributions add a further penalty to the structure, yielding $\Delta F_{300K}$=8.8\,kJ/mol.
In these cases, the maximum number of four backbone H bonds is formed.
In a DFT study of Ac-Ala-Pro-NMe, Byun \textit{et al.} also predicted a $\beta VI$-turn as the most stable conformer in the gas phase.\cite{byun2010conformational}
A comparable $\beta II$' turn  was not among the lowest-energy conformers of this shorter peptide.
The lowest minima of the PES of ADPA (up to 0.7\,kJ/mol) are again $\beta VI$ turns (Figure~3), followed by two alternative conformers with relative potential energies of 2 and 4\,kJ/mol, respectively.
In the latter, the Asp sidechain forms H bonds to the \ce{NH} and \ce{C=O} groups of residue Ala4, resembling the shape of a $\beta$-turn, hence the naming as SC-$\beta$.
For the ADPA-cation systems, we confirmed by mass spectrometry (for the ADPA-cation systems) that the Asp sidechain is protonated in our experimental setup.  
Consequently, the Asp sidechain is modeled in the protonated neutral state.
Harmonic free energy contributions make SC-$\beta$ the preferred structure type by roughly 2\,kJ/mol.
Notably, the lowest free energy structure of AAPA features a \textit{cis} prolyl-peptide bond, while the respective bond in the lowest free energy structure of ADPA is \textit{trans} configured (Figures~2 and 3).

\begin{figure}
 \includegraphics[width=1\textwidth]{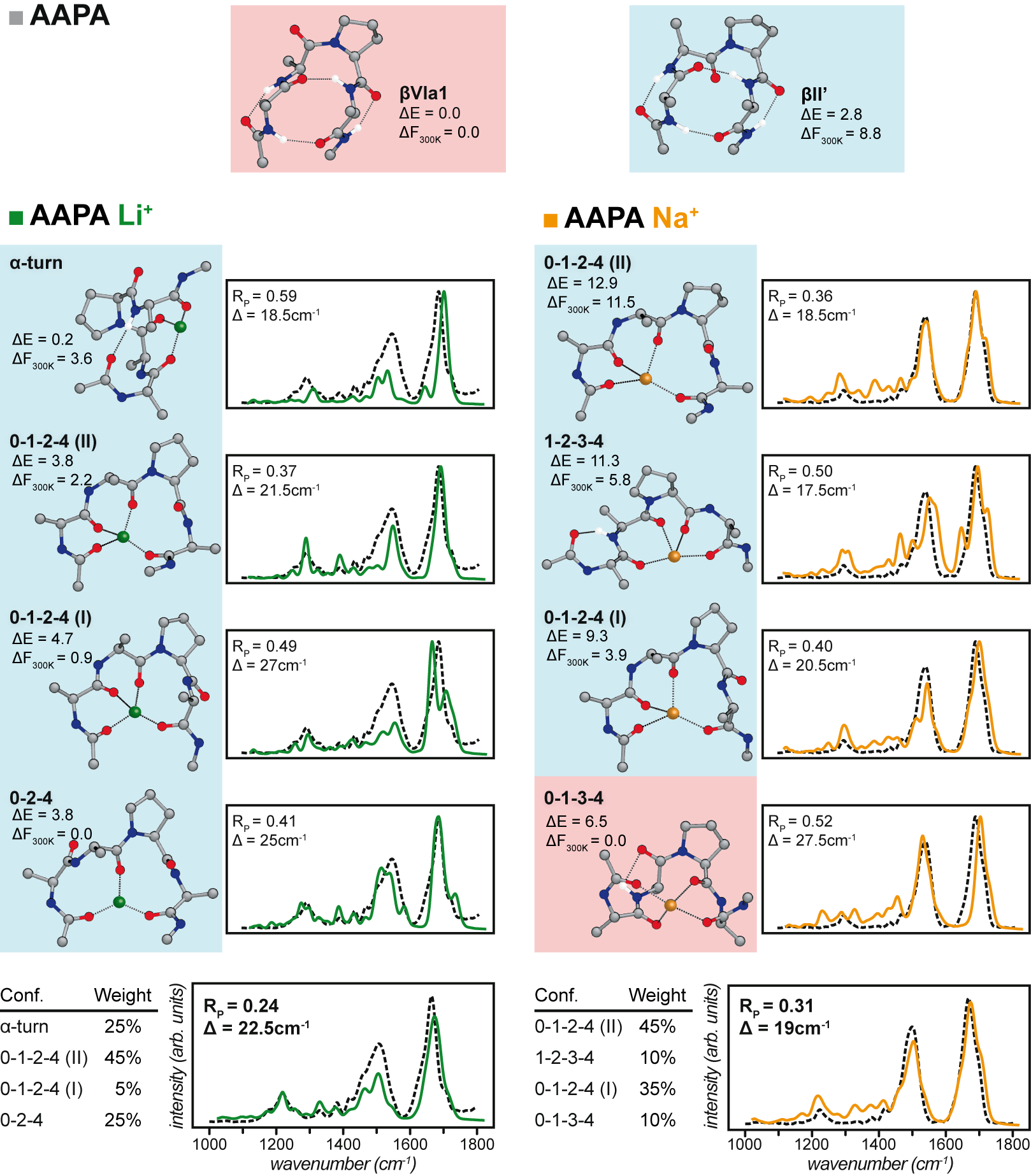}
 \caption{
 \footnotesize
By an exhaustive conformational search we predicted the low free energy ensembles for AAPA in isolation and in the presence of \ce{Li+} and \ce{Na+}. 
The criterion applied to select the structures shown is lowest free energy, except for conformer 0-1-2-4(II) which was selected for its relation to 0-1-2-4(I): both conformers can be interconverted by a backbone crankshaft movement.
It is noteworthy how potential energy ($\Delta E$, in kJ/mol) and harmonic free energy ($\Delta F_{300K}$, in kJ/mol) hierarchies differ for the peptide ion complexes. 
The preference for \textit{cis} (red background) and \textit{trans} (blue background) depends on the cation complexated by AAPA. 
The simulated IR spectra are shown as continuous lines for the individual conformers as well as for the assumed ensemble of conformers (lowest row), always in comparison to the experimental IR spectra in dashed lines.
The tables show the relative proportion of each conformer within the respective mixed simulated spectra.  
Simulated spectra were shifted along the energy axis by a value $\Delta$ for an optimal Pendry reliability factor $R_P$. 
The atom colors: C is gray, N is blue, O is red, H is white, Li is green, and Na is orange. Hydrogens are only shown if part of H bonds.  
}
 \label{fig:AAPA}
\end{figure}

\begin{figure}
 \includegraphics[width=1\textwidth]{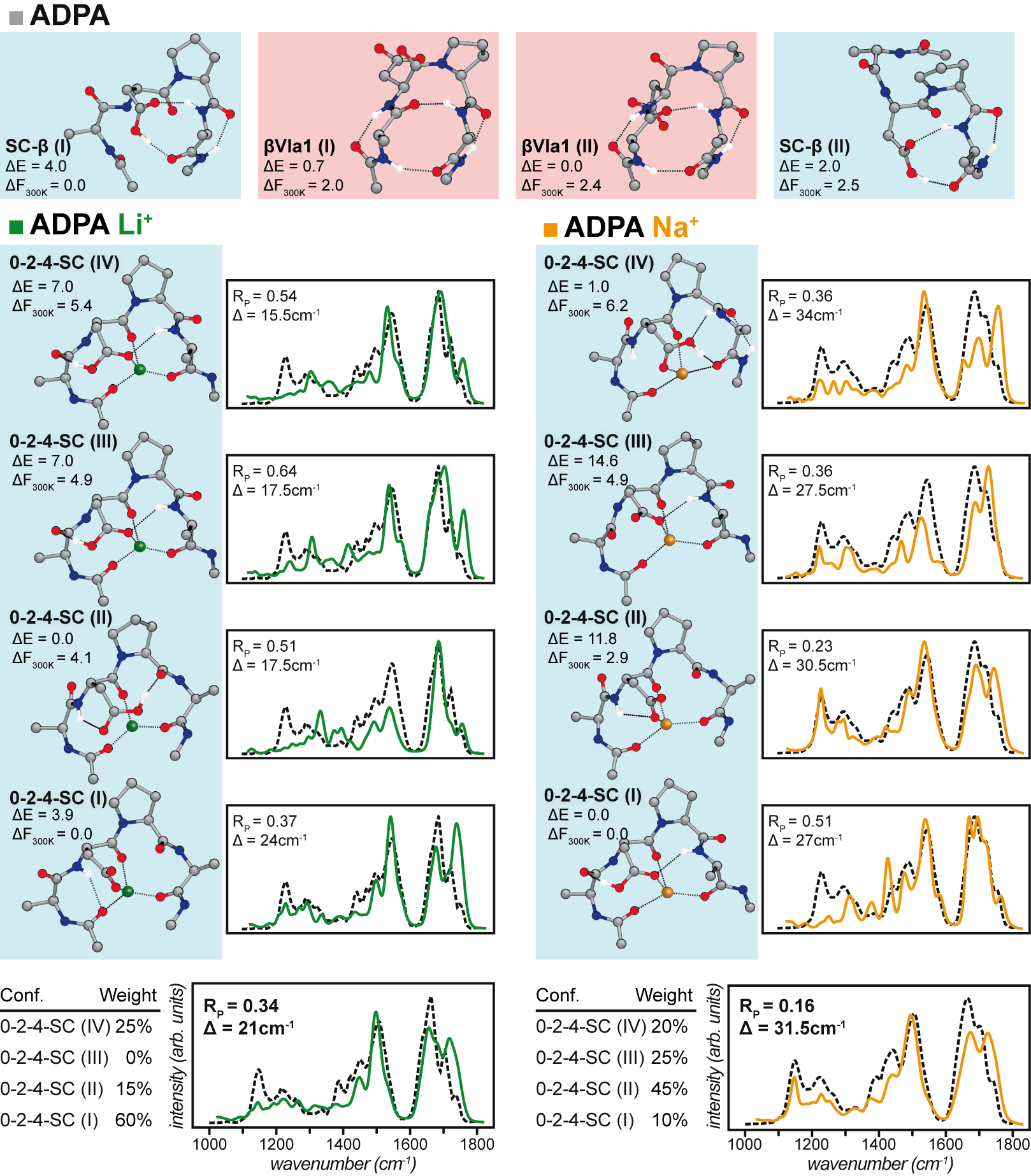}
 \caption{
 \footnotesize
Low free energy conformers for peptide ADPA in isolation and in the presence of \ce{Li+} and \ce{Na+}.
Potential energy ($\Delta E$, in kJ/mol) and harmonic free energy ($\Delta F_{300K}$, in kJ/mol) are given.
The prolyl peptide bond of the respective lowest free energy conformer is \textit{trans} (blue background) for all three cases.
The conformational ensembles of ADPA with \ce{Li+} or \ce{Na+} are very similar, which is also indicated by the similarity of the experimental IR spectra (dashed lines).
Bottom: The simulated spectra (continuous lines) were mixed to account for a conformational ensemble, the tables show the relative proportion of the conformers within the mixed spectra. 
Simulated spectra were shifted by a value $\Delta$ along the energy axis for an optimal Pendry reliability factor $R_P$.
The atom colors: C is gray, N is blue, O is red, H is white, Li is green, and Na is orange. Hydrogens are only shown if part of H bonds. 
}
 \label{fig:ADPA}
\end{figure}

The attraction between backbone carbonyl groups and the \ce{Li+} or \ce{Na+} cations induces structures that differ substantially from the conformers without cations: The H bonding networks in the low energy conformers are disrupted (Figures~2 and 3) and the backbone conformations deviate from the isolated case.
This is in line with the above mentioned results for CysA in apolar Li salt solutions\cite{kessler1990complexation,kock1992novel} and the sodiated peptide-based catalyst in the gas phase.\cite{Garand10022012}
For the isolated peptides AAPA and ADPA, the backbone torsion angles $\phi$ and $\psi$ of the low free energy conformers ($\Delta F_{300K}\,<$\,6\,kJ/mol) are within the allowed regions of the Ramachandran plot (Figure~4).
The single outlier for ADPA in the fourth quadrant represents the C terminal residue Ala4 of a conformer with $\Delta F_{300K}\,=$\,4.8\,kJ/mol.
The different possible rotameric states of the Asp sidechain prefer different backbone conformations.
This results in more possible backbone conformations (data points) compared to AAPA.
The cation-peptide interaction imprints $\phi$/$\psi$ combinations (backbone conformations) that substantially differ from those of the unperturbed peptides. 
Some of them with still low relative free energy (0.9\,kJ/mol for AAPA + \ce{Li+} to 2.6\,kJ/mol for AAPA + \ce{Na+}) are even located outside of the allowed regions of the Ramachandran plot (Figure~4).
These outliers are not at the termini but in the central residues Ala2 or Asp2, which govern the overall structure of the peptides.
Interestingly, also the cation effects on the two peptides differ.
It becomes obvious that the conformational ensembles of AAPA with \ce{Li+} and \ce{Na+}, respectively, differ (Figures~2 and 4), while for lithiated or sodiated ADPA the possible conformations seem very similar (Figures~3 and 4).

A canonical turn structure, type $\beta II$' (not shown), is the lowest PES minimum of AAPA+\ce{Li+}.
The second most stable minimum, with $\Delta E$=\,0.2\,kJ/mol, is a $\alpha$-turn (Figure~2).
Here, the consideration of harmonic free energy contributions changes the picture dramatically and 'unusual' backbone conformations become dominant.
In the lowest free energy conformer, the \ce{Li+} cation is coordinated by three backbone carbonyl groups of residues 0, 2, and 4 (Figure~2).
We base the naming of the conformers on the peptide cation interaction by using the numbers of the interacting oxygens, e.g. 0-2-4.
In case of multiple conformations with the same interaction pattern, they are distinguished with roman numerals, increasing with the free energy of the conformers.
Up to four out of five possible binding partners (backbone carbonyl groups) are sterically possible (conformers 0-1-2-4 with $\Delta F_{300K}$=0.9 or 2.2\,kJ/mol). 
Although the search for minima does not yield information on the actual barriers connecting different conformers, their high structural similarity suggests dynamic interconversion at finite temperature.
For AAPA+\ce{Li+}, the preferred conformation of the prolyl-peptide bond changes from \textit{cis} to \textit{trans}.

\ce{Na+} binding to AAPA results in a similar behavior: Canonical structure types ($\beta VI$, $\beta II$', $\alpha$) are lowest in potential energy while structures with 'unusual' backbone conformations and carbonyl groups pointing towards \ce{Na+} are most stable when harmonic free energy contributions are considered.
However, there are substantial differences to the \ce{Li+} adducts: the low free energy ensemble is more diverse and the central peptide bond of the lowest free energy conformer 0-1-3-4 is \textit{cis}.
In the case of AAPA + \ce{Na+} (Figure~2), we do not show just the four lowest free-energy conformers (see Table 3 of the Supporting Information for all calculated free energies). 
The second lowest free energy conformer, A1661 ($\Delta F_{300K}$=2.6\,kJ/mol), was excluded since it was proven unstable in the subsequent AIMD simulations for IR spectra (see section below).
Instead we consider 0-1-2-4(II) as fourth conformer, which is much higher in free energy.
Interestingly, these two conformers of AAPA+\ce{Na+}, 0-1-2-4(I) and 0-1-2-4(II), are almost identical besides the orientation peptide bond between Pro3 and Ala4 (Figure~2).
This peptide bond is not involved in any interactions and can thus rotate by a concerted motion of the adjacent torsion angles $\psi$ and $\phi$, a so-called backbone crankshaft rotation.\cite{wasserman1990elastin,fadel1995crankshaft}
During the equilibration AIMD simulations at 300\,K (in preparation of the simulations to obtain IR spectra), this interconversion from 0-1-2-4(I) to 0-1-2-4(II) was indeed observed within the 10\,ps simulation time. 
The subsequent evaluation of IR spectra also suggests the presence of 0-1-2-4(II) in the experimentally observed conformational ensemble.

In ADPA, the dominant interaction pattern is the complexation of \ce{Li+} or \ce{Na+} by the backbone oxygens 0, 2, 4 and the Asp sidechain carboxyl group.
All conformers in the low energy range are highly similar and feature no \textit{cis} prolyl peptide bonds (Figure~3).
As discussed above on the basis of the Ramachandran plot (Figure~4), the effects of the cations on AAPA and ADPA differ.
\ce{Li+} enforces a \textit{trans} conformation on the prolyl peptide bond of AAPA while \ce{Na+} enforces the \textit{cis} conformation (Figure~2).
For ADPA, no such selectivity for the cation is observed.
With \ce{Li+} or \ce{Na+} attached, similar structure types with \textit{trans} prolyl peptide bonds are preferred (Figures~3 and 4).

\begin{figure}
 \includegraphics[width=0.48\textwidth]{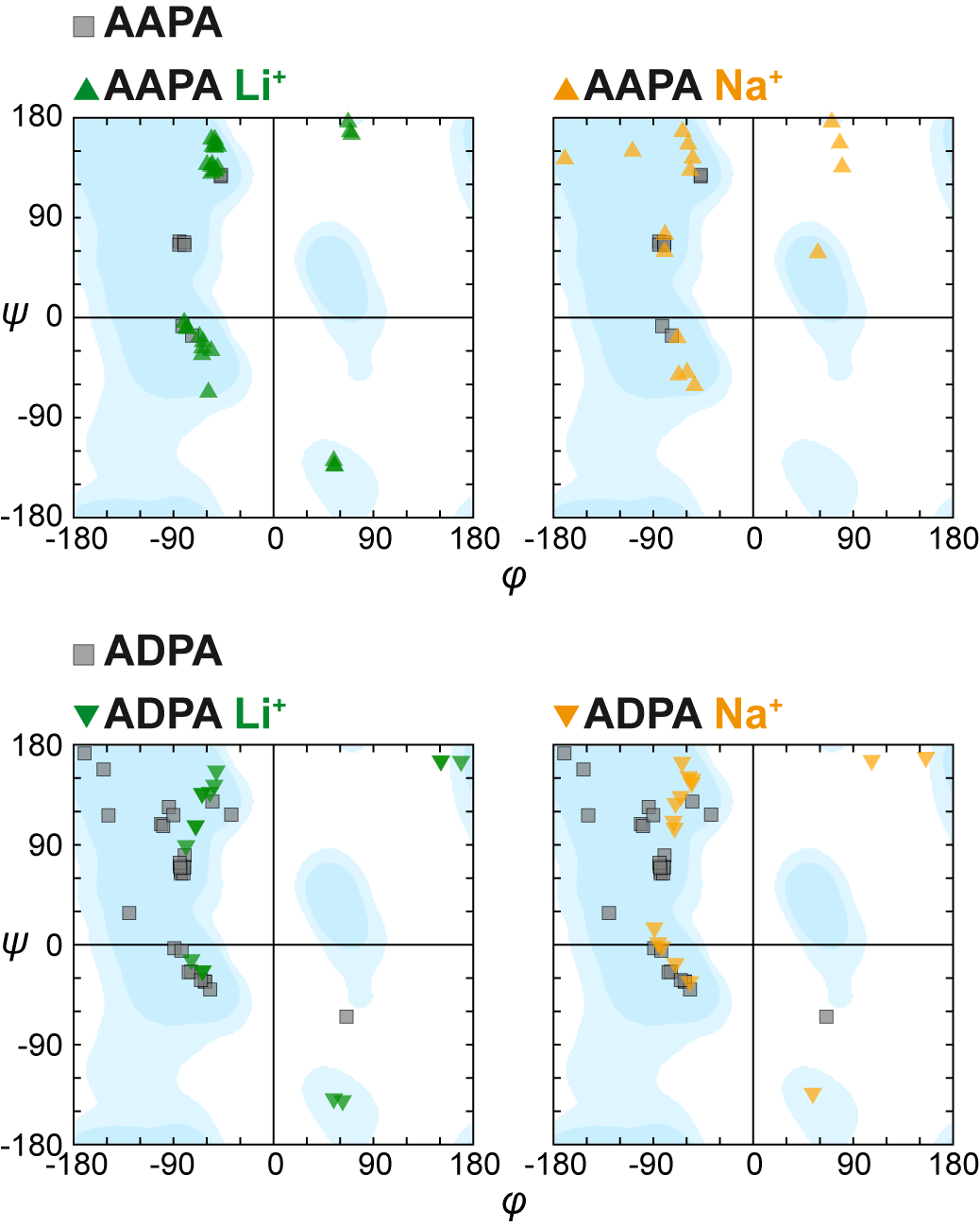}
 \caption{The backbone torsion angles $\phi$ and $\psi$ of the low free energy conformers ($\Delta F_{300K}<\,6$\,kJ/mol) for AAPA and ADPA in isolation (gray squares), with \ce{Li+} (green triangles), and with \ce{Na+} (orange triangles) were plotted on top of an empiric contour plot.\cite{lovell2003structurevalidation} 
 Addition of cations disturbs the backbone conformations and even results in usually forbidden $\phi$/$\psi$.
 The backbone conformation of AAPA is sensitive to the type of cation, as illustrated by the differing torsion angle patterns with \ce{Li+} and \ce{Na+}.
}
 \label{fig:Rama_Plots}
\end{figure}
 
\subsection{Infrared spectroscopy}
In order to corroborate our structural findings, we obtained gas-phase infrared multi-photon dissociation (IRMPD) spectra, which reflect the same clean-room conditions as used in our simulations.
Spectra were recorded from 1000 to 1800\,cm$^{-1}$ at the free electron laser facility FELIX \cite{oepts1995free} using a Fourier-transform ion cyclotron (FT-ICR) mass spectrometer.\cite{valle2005free}
The experimentally obtained spectra for lithiated and sodiated AAPA and ADPA are shown in Figure~2 and Figure~3.
For AAPA significantly different spectral signatures where obtained for the \ce{Li+} and \ce{Na+} complexed forms, which is in line with the results of the coformational analysis described in the previous section. 
For ADPA on the other hand very similar spectra were recorded with both cations.

In order to allow for a quantitative theory-experiment comparison, IR spectra including anharmonic effects were computed from Born-Oppenheimer \textit{ab initio} molecular dynamics (AIMD) simulations.  
The systems were equilibrated by 10\,ps of AIMD simulations at 300\,K.
Subsequently, the micro-canonical ensemble was sampled in up to 40\,ps long AIMD simulations at constant energy from which IR spectra were derived.\cite{gaigeot2010theoretical,rossi2010secondary}
IR spectra of polyamides feature characteristic bands of high intensity (like the amide I and II region, 1400\,-\,1700\,cm$^{-1}$) but also regions with low intensity (below 1400\,cm$^{-1}$) and fingerprint characteristics. 
Visual inspection does not allow for a quantitative assessment and is, similar to a simple square of intensity comparison, easily biased by the high intensity peaks.
For a quantitative comparison between the calculated and experimental spectra, we employ the reliability factor $R_P$ introduced by Pendry \cite{pendry1980reliability} to the field of low-energy electron diffraction, in an implementation distributed with reference~\onlinecite{blum2001fast}.
For $R_P$, peak positions are more important than peak intensities -- a characteristic that fits the requirements we face here, especially as we compare experimental action spectra and theoretical absorption spectra.
Values for $R_P$ range from 0 (perfect agreement) to 1 (no correlation).
Intensities of the spectra were normalized to 1 and rigidly shifted (not scaled) with a value $\Delta$ along the energy axis to account for deviations likely due to a systematic mode-softening by the density functional we use.\cite{gregoire2007resonant,rossi2010secondary}
When comparing the calculated IR spectra of single conformations to the experimental IR spectra we observe only modest agreement (see individual spectra in Figures~2 and 3).
Previous studies have shown similar behavior due to conformational ensembles for peptides in the gas-phase at finite temperature.\cite{compagnon2006peptides,kamariotis2006infrared,cimas2009vibrational,james2009intramol,james2010laserspec,rossi2010secondary}
Furthermore, the energy differences of the low free-energy conformers lie within the uncertainty of the employed method, as discussed in the benchmarks section above.
Consequently, an ensemble of conformations is assumed.
By mixing the individual theoretical spectra in 5\% steps, the $R_P$ to the respective experimental spectrum is optimized.
This results in a much better agreement of simulated and experimental spectra of the peptides AAPA and ADPA in complex with single \ce{Li+} or \ce{Na+} cations (Figures~2 and 3).
For the predicted spectra of AAPA + \ce{Li+} and for ADPA + \ce{Li+} or \ce{Na+}, especially the reproduction of the fine structure below 1400\,cm$^{-1}$ wavenumbers is gratifying.
We note for completeness that the spectra for the protonated peptides (not shown) are rather different in appearance, suggesting very different structural effects compared to the heavier cations.

In a naive way, a correlation between the free energy estimates at the harmonic approximation and the abundances of the individual spectra in the resulting mixed spectrum could be expected. 
However, this would be too much to expect for several reasons: 
\begin{itemize}
  \item The PBE+vdW method we use is rather accurate as illustrated by the benchmark calculation presented in the methodology section of this article, however, the systems under investigation here are also large (56 to 60 atoms). The lowest free-energy minima we discuss here are still within the possible uncertainties of the relative (free) energies.
  \item  The experimental data base to which we are comparing mixes of theoretical spectra is, simply, small - and fitting many parameters to a small data set has well known limitations.\cite{pendry1980reliability}  What these mixes offer is, therefore, strictly only a consistency check. The spectra of just a single conformer are not sufficient to explain the observed IR spectra. In contrast, conformational mixes yield a much more consistent description of the spectra, in line with several conformers of similar free energy. This is the primary quantitative statement that we can derive from the experiment-theory comparison.
  \item The free energy model neglects anharmonicity as well as the entropic effects of a possibly greater accessible conformational space (dynamical interconversion in the case of low barriers) of specific conformers. That the latter can be of special importance is illustrated by the crankshaft rotation discussed above for the two conformers 0-1-2-4(I) and 0-1-2-4(II) of AAPA + \ce{Na+}. The fact that both conformers can interconvert shows the extent to which the anharmonic nature of the potential-energy surface can play a role. In fact, in the combination of the four individual spectra that shows the best agreement with the experimental spectrum (Figure~2), 0-1-2-4(II) is predominant with 45\%. Again, conformer 0-1-2-4(II) is structurally and dynamically closely related to the 0-1-2-4(I) conformer with the lowest harmonic free energy. This illustrates the limits of the harmonic free energy assignment to potential energy minima at room temperature, where such conformational and dynamical effects are neglected. It furthermore illustrates the limitations of the interpretation of IR spectra as a combination of individual and isolated conformers.
\end{itemize}

Concluding this section we can say that, on the one hand, the accuracy of the harmonic approximation to the free energy is limited by the dynamic character of such molecular systems at finite T.
On the other, the IRMPD spectroscopy setup we use here is limited in its resolution, especially regarding the separation of individual conformers.
However, we can unambiguously predict minima by first-principles theory and validate the results by room temperature IR spectroscopy (keeping the differences of static harmonic free-energy minima and actual room temperature molecules in mind).
The observed cation-peptide effects were certainly qualitatively corroborated by both approaches.

\subsection{Micro solvation of a peptide cation complex}
Already in the introduction we mention the presence of turn-sequences mainly at the surface of proteins, exposed to the aqueous environment.
With this section we want to give at least a qualitative picture of how the interaction between the peptide backbone and the cation can compete with solvation of the cation.
AIMD simulations have been performed for AAPA+\ce{Li+} alone and with a few water molecules.
For the setup of the latter system, 18 molecules were accommodated within a sphere of 4.5\,\AA{} radius around the \ce{Li+} cation.
For comparison, \ce{Li+} embedded in 4 and 10 water molecules, respectively, was also studied.
We characterize the interaction between the \ce{Li+} cation and the respective oxygens of the peptide backbone or of first solvation shell water molecules by the \ce{Li+}-O distance and by the O-\ce{Li+}-O angle, shown in Figure~5.
Previous \textit{ab initio} studies predict a coordination number of 4 for \ce{Li+} in water.\cite{kamariotis2006infrared,varma2006,ikeda:034501} 
Consistently, the cation is complexed by 4 backbone carbonyl groups in our example conformer 0-1-2-4(I).
During a 100\,ps AIMD trajectory at 330\,K (Nos\'{e}-Hoover thermostat), the \ce{Li+}-O distance fluctuates around 1.9\,\AA{}, the O-\ce{Li+}-O angle distribution is broad, indicating the non-ideal tetrahedron formed by the interacting carbonyl oxygens.
The microsolvation of AAPA + \ce{Li+} in 18 water molecules results in a slight change of the binding site within a few picoseconds: a water oxygen substitutes for the interaction to backbone \ce{C=O} of Ala1.
The cation interacts with the three backbone carbonyl oxygens of AAPA and the same water molecule (Figure~5) for the whole 90\,ps of remaining AIMD simulation time.
As a result, a virtually ideal binding site is formed, characterized by an almost symmetric distribution of the O-\ce{Li+}-O angle around the ideal tetrahedral angle of 109.5$^\circ$.
For \ce{Li+} immersed in a small water cluster (4 or 10 water molecules, respectively), the \ce{Li+}-O distance distribution peaks around 2.0\,\AA{}. 
Remarkably, the distribution of the tetrahedron angles O-\ce{Li+}-O is multimodal again, accounting for alternative (and less populated) geometries of the \ce{Li+} complexation involving 3 or, in the case of the 10 water with \ce{Li+} cluster, even 5 water molecules in the first solvation shell.
For now, we can at least qualitatively say that AAPA is able to form an ideal interaction shell that seems competitive to water solvation.
A fully correct answer could be given on the basis of free energy differences from simulations with fully solvated systems.
Such simulations are standard for force field approaches, yet they are computationally very demanding at the level of theory we employ here.
A rigorous assessment is thus beyond the scope of this manuscript.

\begin{figure}
 \includegraphics[width=1\textwidth]{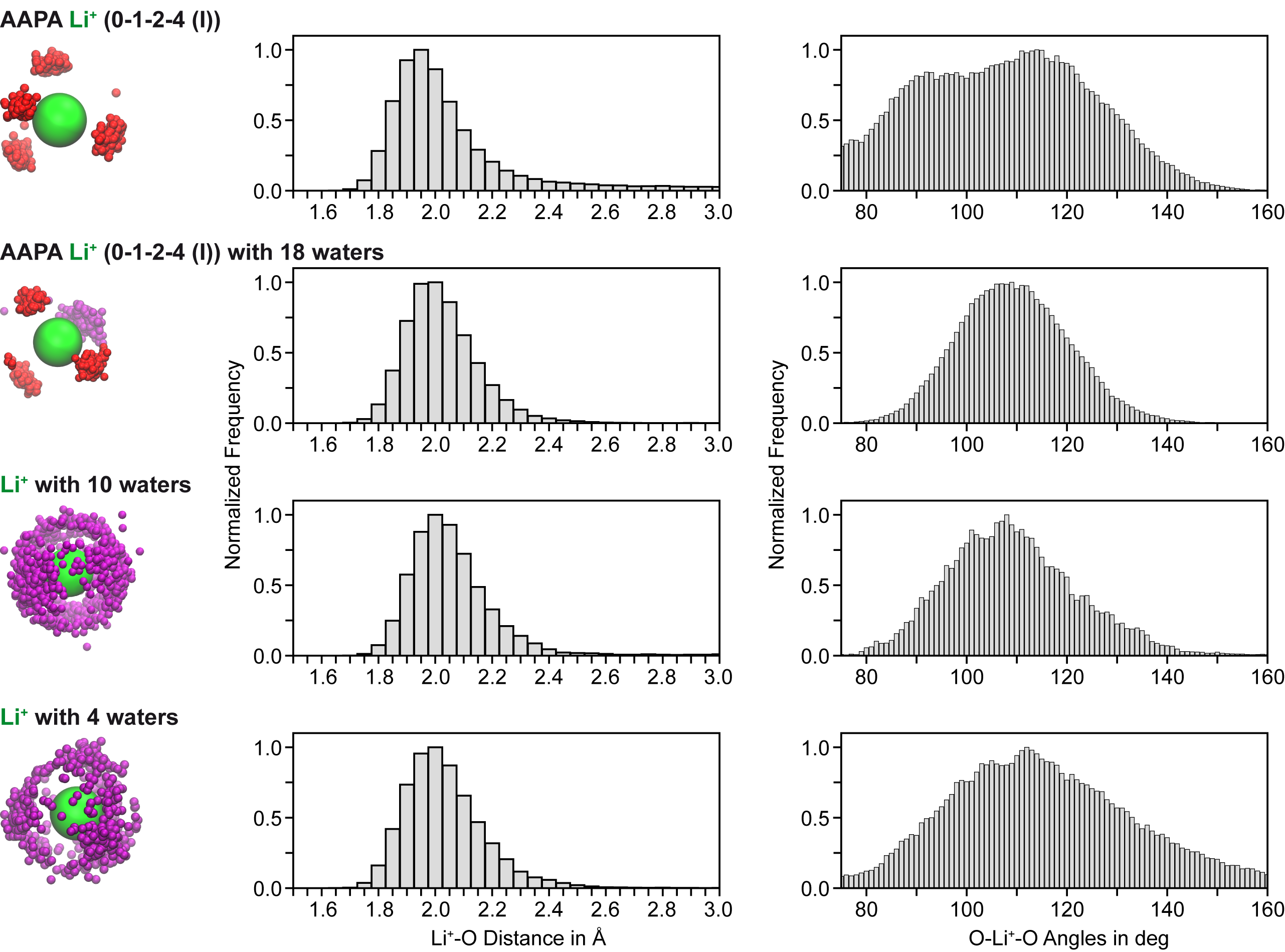}
 \caption{The \textit{first solvation shell} around a \ce{Li+} cation. Interactions are formed between between the oxygens of water molecules (purple) or backbone carbonyl groups (red). The histograms were derived from AIMD simulation of different length (20 to 100\,ps) and the counts were normalized to 1.
  }
 \label{fig:firstsolvationshell}
\end{figure}

\subsection{Conclusion}
Starting from the isolated peptides that adopt either canonical turn structures (AAPA) or turn-like conformations with sidechain to backbone H bonds (ADPA), we show the drastic effect of cations on the local secondary structure of peptides: the cation attracts most of the backbone carbonyl groups and, as a result, completely breaks the local H bonding network. 
This leads to distortions of the peptide backbone and results in conformations with backbone torsion angles $\phi$ and $\psi$ that are in part outside of the allowed regions of the Ramachandran plot (Figure~4).
Consequently the question of the range of such ion-induced disruptions arises.
Ohanessian and co-workers studied \cite{Semrouni2010728,balaj2012sodiated} poly-glycines with a chain length of 2 to 8 in complex with \ce{Na+} by simulation and gas-phase IR spectroscopy:
Up to a sequence of 7 glycine residues the contact number between the cation and backbone \ce{C=O} groups is maximized and no H bonding was observed.
With the Gly$_8$ peptide, backbone H bonding appeared again in form of $\gamma$- and $\beta$- turns.
Glycine, due to the lack of a sidechain, represents a very special case among the canonical amino acids.
As a contrast, the helical secondary structure of sodiated polyalanine (8-12) is not broken in the gas phase. 
Here, the \ce{Na+} ion is attached to the C terminus.\cite{kohtani2000metal,doi:10.1021/jz301326w} 
The importance of considering the effect of sidechain functionalities is highlighted by the sequence dependence of the cation effects we observe.
The conformational preferences of AAPA with \ce{Li+} or \ce{Na+} differ drastically by the \textit{trans} or \textit{cis} state of the central prolyl peptide bond.
Noskov and Roux investigated the selectivity of the ion-coupled transporter LeuT.
Two \ce{Na+} binding sites (NA1 and NA2) show differences in the \ce{Li+}/\ce{Na+} selectivity: NA1 appears to be rather flexible and exhibits no selectivity for one cation over the other as it adapts to the different ionic radii.
For NA2, a slight selectivity is apparently induced by a ''snug-fit`` mechanism (the rigid NA2 interaction site is unable to adapt to different ionic radii).\cite{Noskov2008804}
Similarly, the lowest free-energy structure 0-1-3-4 of AAPA+\ce{Na+} may be to rigid to adapt to the \ce{Li+} cation, since only backbone carbonyl functions can be involved in the interaction.
With ADPA, the Asp sidechain prevents such conformation selectivity.

Our findings might even help to understand a basic biochemical principle: In 1888, Hofmeister published an article \cite{hofmeister1888salze} that laid the basis for a sorting of cations and anions according to their effect on the solubility of biomolecules, colloids, and functional polymers.
While it was believed that the underlying effects can be explained solely by bulk properties stemming from the solvent ion interactions,\cite{vonHippel1965confstab} evidence was found that most effects of ions on water structure are limited to the first solvation shell\cite{Omta18072003} and that specific ion-solute interactions can be expected to contribute substantially.\cite{kunz2010specific}
This gets especially clear at high salt concentrations as investigated by Dzubiella and co-workers employing classical MD simulations\cite{dzubiella2008salt,dzubiella2009salt,vonhansen2010ionspeci} and later also experimental approaches.\cite{crevenna2012hofmeister}
They demonstrate that the perturbing effect of ions on peptide structure results from the breaking secondary structure specific H bonds in the backbone.
Our own findings point into a similar direction, as we have shown here how cations can substantially change the backbone structure of a (bio)polymer.
These interactions are not necessarily stable over a very long time range, but our exploratory AIMD simulations suggest time ranges at least in the tens to hundreds picoseconds.
Dzubiella describes long-lived loop conformations that are stable over 10 to 20\,ns in classical MD trajectories.\cite{dzubiella2008salt}
Similar to the \emph{specific interactions} between anions and the amide bond containing polymers of N-isopropylacrylamide described by Cremer and co-workers,\cite{zhang2005specific} we show here the possible interactions between small mono-valent cations and peptides and highlight their significant effect on local peptide structure.
Such should be considered as one driver behind the Hofmeister salt effects on proteins.

\section{Computational methods}
 
Scans of the PES were peformed with a basin hopping-like exhaustive search and conventional protein force fields (OPLS-AA \cite{jorgensen2004free} or AMBER99 \cite{wang2000well}). 
We employ the TINKER~5 scan routine \cite{pappu1998analysis} in an in-house parallelized version.
The required methods to perform DFT-based simulations, including geometry optimization, computation of harmonic vibrations, and \textit{ab initio} Born-Oppenheimer molecular dynamics (AIMD), are incorporated in the FHI-aims code, which provides an efficient and accurate all-electron description based on numeric atom-centered orbitals.\cite{blum2009ab} 
In the following we discuss fully relaxed conformations at the PBE+vdW level and their relative potential energies ($\Delta E$) and relative harmonic free energies at 300K ($\Delta F_{300K}$), all computed with \textsl{tight} convergence settings and an accurate \textsl{tier~2} basis set.\cite{blum2009ab} 
High-level quantum chemical benchmark calculations, i.e. relaxations at the MP2 level of theory and coupled-cluster calculations with singles, doubles and perturbative triples (CCSD(T)), were performed with the ORCA quantum chemistry program,\cite{Orca2012}
CCSD(T) energies extrapolated to the complete basis set limit (CBS) were obtained by a method described by Truhlar,\cite{Truhlar1998extrapolation} employing the Dunning basis sets cc-pVDZ and cc-pVTZ.\cite{dunning1989basis}

\subsection{Benchmarks}

We assessed the predictive power of the DFT approximations applied here by benchmarks in two directions with respect to the approximation level:
We compare to high-level quantum chemistry at the CCSD(T) level of theory extrapolated to the complete basis-set limit.
On the other hand, we assess the quality of the force field description of cation-peptide interactions in comparison to approximate DFT at the PBE+vdW and PBE0+vdW levels.

\subsubsection{Comparing electronic structure theory methods}
There have been several assessments of the accuracy of the PBE+vdW level of theory applied to a variety of systems, e.g., peptides,\cite{PhysRevLett.106.118102} weakly bound metal-phtalocyanine systems,\cite{Marom2010} and ionic and semiconductor solids.\cite{zhang2011vdwsolid}
A previous assessment of the accuracy of PBE+vdW for peptide systems, for the conformational energy hierarchy of Ace-Ala-NMe and Ace-Ala$_3$-NMe, shows mean absolute errors (MAE) below 2\,kJ/mol in comparison to CCSD(T) energies.\cite{PhysRevLett.106.118102}
We here investigate metal cation-peptide systems and thus re-assess the accuracy of our DFT-based predictions.
We employ high-level quantum chemical theory on the conformational energy hierarchy of Ac-Ala-NMe+\ce{Li+}.
A conformational analysis identified five local minima (Figure~6) within a potential energy range of 35\,kJ/mol at the MP2/cc-pVTZ level of theory \cite{pople1976mp2,dunning1989basis}.
The cation closes a 7-membered pseudo-cycle via interaction with the oxygens of the backbone carbonyl groups. 
The orientation of the methyl groups relative to the pseudo-ring plane defines them as either equatorial (Figure~6: \textbf{1}, \textbf{3}, \textbf{5}) or axial (Figure~6: \textbf{2}, \textbf{4}).
In addition, an important characteristic of our actual systems of interest (AAPA and ADPA) is present here as well: in the lower energy conformers \textbf{1} and \textbf{2} the C-terminal peptide bond is \textit{trans} configured, in contrast to the other conformers with a C terminal \textit{cis} peptide bond.

\begin{figure}
 \includegraphics[width=0.48\textwidth]{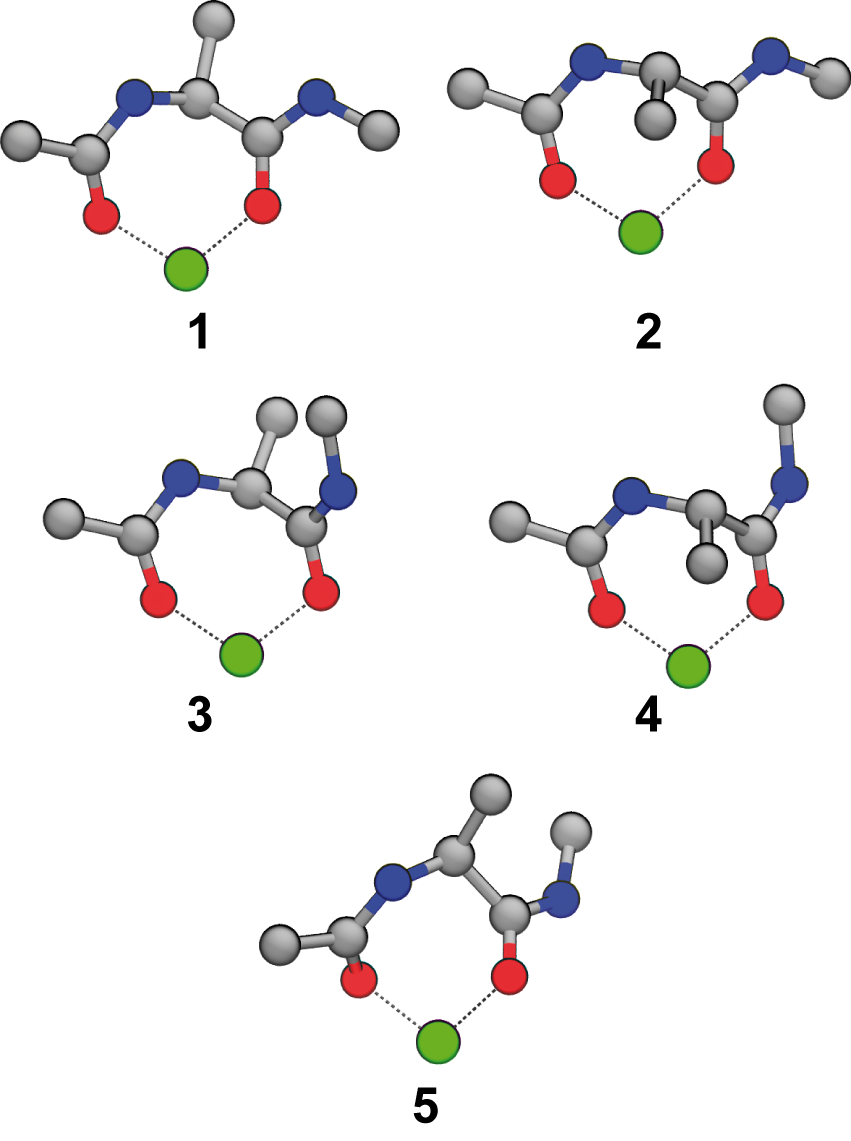}
 \caption{Lowest energy conformers of Ace-Ala-NHMe + \ce{Li+}, fully relaxed at the MP2/cc-pVTZ level of \textit{ab initio} theory. Hydrogens were omitted for clarity; dashed black lines illustrate the oxygen lithium interactions.
}
 \label{fig:AlaLi_confs}
\end{figure}

Relative energies at the CCSD(T), PBE+vdW, and PBE0+vdW level of theory were compared by estimating the mean absolute error ($MAE= \frac{1}{n}  \sum_{i=1}^{n}|f_i - y_i|$).
As an additional method, we also include the frequently-used B3LYP method (no vdW correction).
For much larger systems, the missing description of dispersion effects represents a deficiency in the description of conformational energy hierarchies.
The DFT methods give low uncertainties (see Table~1), well within the often stressed ''chemical accuracy`` of 1\,kcal/mol (4.2\,kJ/mol).
The conformational changes upon relaxation with DFT are negligible, as indicated by the low maximal RMSD value of 0.38\,\AA{} (Table~1).
We note that only small contributions of the van der Waals correction can be expected for molecular systems of this size.
Of the tested DFT approaches, PBE0+vdW gives the best agreement with the benchmark calculations, as it is obvious from the MAE and the average RMSD (see Table~1).
However, for a large-scale conformational screening and the extensive molecular dynamics simulations we undertake in this study, PBE+vdW perfectly balances computational costs and accuracy. 

\begin{table}
  \caption{Relative energies and RMSD values of the conformers depicted in Figure~Figure~6. 
  Left columns: CCSD(T), PBE+vdW, PBE0+vdW, and B3LYP relative energies were calculated for MP2/cc-pVTZ geometries.
  The MAE of the DFT relative energy hierarchies to CCSD(T) is also given. 
  Right columns: the conformers were also relaxed with the respective DFT methods.
  With respect to the MP2 geometries, RMSD values for the individual conformers and average  RMSD values are given.
  Relative energies and the mean absolute errors (MAE) are given in kJ/mol; RMSD values are given in \AA{}. 
  }
  \label{tbl:energies}
  \begin{tabular}{l|rrrr|rrr}
    \hline
    & \mc{4}{c|}{$\Delta E$ (MP2 geometries)} & \mc{3}{c}{RMSD to MP2} \\
Conf.     & CCSD(T) & PBE+vdW & PBE0+vdW & B3LYP  & PBE+vdW & PBE0+vdW & B3LYP  \\
    \hline
1 &	0.0 & 0.0 & 0.0 & 0.0 &      0.03 & 0.03 & 0.03  \\
2 &	6.7 & 4.4 & 5.0 & 5.7 &      0.03 & 0.03 & 0.03  \\
3 &	20.2 & 17.1 & 18.2 & 22.3 &  0.38 & 0.16 & 0.33  \\
4 &	23.1 & 22.0 & 23.1 & 27.1 &  0.12 & 0.08 & 0.18  \\
5 &	33.8 & 34.2 & 34.8 & 38.0 &  0.04 & 0.02 & 0.04  \\
\hline
\mc{2}{l}{MAE/RMSD} & 1.4 & 0.9 & 2.3 & 0.12 & 0.06 & 0.12  \\
\hline
\end{tabular}
\end{table}

\subsubsection{Standard protein force fields versus electronic structure theory}
When comparing the results of different standard protein force fields among each other and to DFT, we observe dramatic discrepancies in the conformational hierarchies.
Such force fields were parametrized for the solvated state, while we perform our assessment in the gas-phase. 
Nonetheless, they are frequently used also for conformational investigations irrespective of the environment.
Consequently, their performance \textit{in vacuo} is of interest.
We employ as a reference the conformational energy hierarchy of AAPA+\ce{Li+} at the PBE+vdW level.
In line with the results of the above comparison, PBE \cite{perdew1996generalized} and the hybrid density functional PBE0 \cite{adamo1999pbe0} (both vdW corrected \cite{tkatchenko2009accurate}) give very similar results, illustrated by the low mean absolute (MAE) and maximal errors (E$_{max}$) listed in Table~2.
Both approaches without the vdW correction give higher MAE and E$_{max}$.
The widely used protein force fields Amber99,\cite{wang2000well} Charmm22,\cite{mackerell1998all} and OPLS-AA \cite{jorgensen2004free} give MAE that are at least about 15 times larger with a dramatic magnitude of the maximal errors (see Table~2).
The relative energies can be found in the Supporting Information.
A main characteristic of the system is apparently the cation peptide interaction.
The effect on the partial charges appears to be better described by the polarizable FF 
Amoeba\cite{schnieders2007polarizable}, illustrated by a MAE of about 10\,kJ/mol.
The removal of the cation leads to reduced MAE values for the FF methods (cf. Table~2), also the energy hierarchies themselves appear more consistent in the different methods.
This can be seen when comparing the two plots (with and without \ce{Li+}) in Figure~7 or when looking at the maximal errors given in Table~2.
The calculations (single point) were repeated for the same AAPA conformers (fixed geometries) but without the cation.
The MAEs to the force field approaches are consistently much larger than between the DFT techniques, with significant errors especially also in the energetic hierarchy of the conformers.
Apparently, the large errors of the force fields can mainly be attributed to the ill-described cation-peptide interaction.
In short, DFT based approaches for cation-peptide systems appear to be vastly preferable at least over the standard force field-based approaches tested here.

\begin{figure}
 \includegraphics[width=0.8\textwidth]{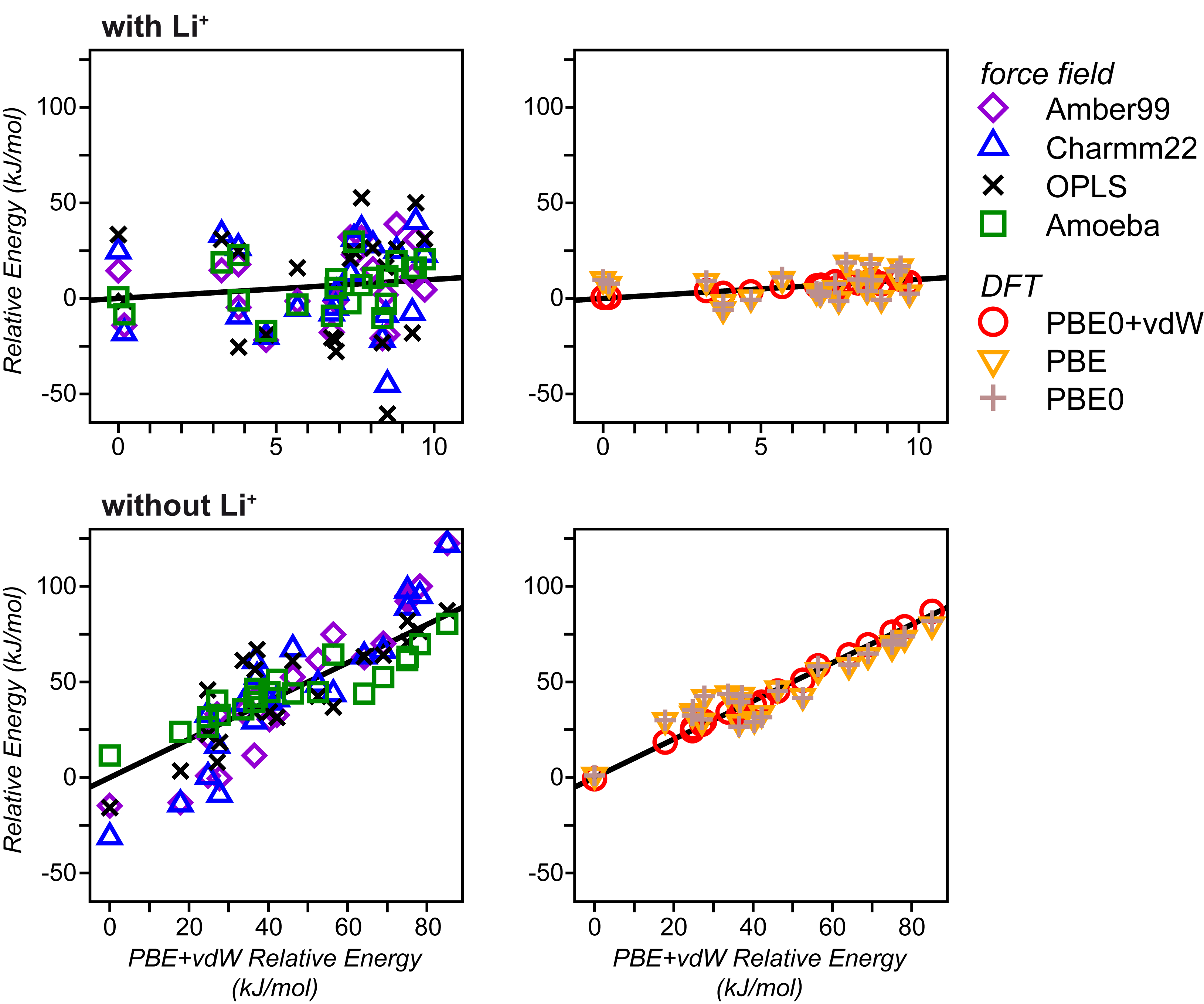}
 \caption{The relative energies of 21 AAPA+\ce{Li+} conformers with a relative potential energy below 10\,kJ/mol at the PBE+vdW level were re-calculated with PBE0+vdW, PBE, PBE0, Amoeba,\cite{schnieders2007polarizable} Amber99,\cite{wang2000well} Charmm22,\cite{mackerell1998all} and OPLS \cite{jorgensen2004free}, for the fixed geometries with and without the \ce{Li+} cation.
 The conformational energy hierarchies with different FF and DFT methods are plotted against the in PBE+vdW values ($x$-axis).
 The relative energy values were shifted according to the overall offset of the individual energy hierarchy.
 Please note the different scales of the $x$-axes in the plots with and without \ce{Li+} cation. 
 The hypothetical perfect correlation is indicated by the straight lines, but all force fields deviate significantly even for the qualitative energetics.
}
 \label{fig:EBench}
\end{figure}

\begin{table}
  \caption{MAE and E$_{max}$, with respect to the PBE+vdW hierarchy, of the energy hierarchies computed with PBE0+vdW, PBE, PBE0, Amoeba,\cite{schnieders2007polarizable} Amber99,\cite{wang2000well} Charmm22,\cite{mackerell1998all} and OPLS \cite{jorgensen2004free}, for the fixed geometries with and without the \ce{Li+} cation (energies in kJ/mol). The relative energies can be found in Tables 1 and 2 of the Supporting Information.
  }
  \label{tbl:FF-MAE}
  \begin{tabular}{lccccccc}
    \hline
    & PBE0+vdW & PBE & PBE0 & Amoeba & Amber99 & Charmm22 & OPLS-AA \\
    \hline
    & \mc{7}{c}{Fixed AAPA+Li$^+$ geometries} \\
  MAE & 1.0 & 5.7 & 6.1 & 9.7 & 15.7 & 20.5 & 26.0 \\
  E$_{max}$ & 2.0 & 10.9 & 11.1 & 22.3 & 30.1 & 53.6 & 69.1 \\
  \hline
    & \mc{7}{c}{Same geometries, fixed without Li$^+$} \\
  MAE & 1.1 & 6.6 & 6.6 & 8.1 & 14.0 & 15.6 & 11.6 \\
  E$_{max}$ & 2.8 & 14.3 & 14.8 & 20.3 & 37.5 & 36.8 & 29.7 \\
    \hline
    \hline
\end{tabular}
\end{table}


\section{Experimental methods}
\subsection{Synthesis}
Peptides were synthesized by solid phase assembly using a Multi-Syntech Syro XP peptide synthesizer (MultisynTech GmbH, Witten, Germany) by Fmoc strategy on Fmoc-Ala-OWang resin (0.5\,mmol/g). 
The peptides were cleaved from the resin by reaction with 2\,ml of a solution containing 10\% (w/v) triisopropylsilane, 1\% (w/v) water, and 89\% (w/v) TFA. 
The crude peptides were purified by reversed-phase HPLC on a Knauer smartline manager 5000 system (Knauer GmbH, Berlin, Germany) equipped with a C8 (10\,$\mu$m) LUNATM Phenomenex column (Phenomenex Inc., Torrance, CA, USA). 
Peptides were eluted with a linear gradient of acetonitrile/water/0.1\% trifluoroacetic acid and identified on an Agilent 6210 ESI-TOF mass spectrometer. 
Peptide purity was determined by analytical HPLC on a Merck LaChrom system (Merck KGaA, Darmstadt, Germany) equipped with a C8 (10\,$\mu$m) LUNATM Phenomenex column (Phenomenex Inc., Torrance, CA, USA). 
The gradient used was similar to those of the preparative HPLC.

\subsection{Infrared spectroscopy}
The gasphase IR experiments were performed at the free electron laser facility FELIX\cite{oepts1995free} (Nieuwegein, The Netherlands) using the Fourier-transform ion cyclotron (FT-ICR) mass spectrometer\cite{valle2005free} which was temporarily equipped with a nano electrospray ionization source (MS Vision, Almere, The Netherlands).
Typically, 5\,$\mu$l of a solution containing 1\,mM peptide, 50\% water, 50\% methanol and, where needed, 10\,mM LiCl or NaCl, were placed in gold-coated, off-line emitters prepared in-house. 
In order to obtain a stable spray, a small backing pressure of approximately 0.5~bar and a relatively low capillary voltage of approximately 850\,V was applied to the needle. 
The nESI generated ions were accumulated in a hexapole ion trap and subsequently transferred into the FT-ICR mass spectrometer that is optically accessible via a KRS-5 window at the back end. 
After trapping and SWIFT mass-isolation inside the ICR cell, the ions were irradiated by IR photons of the free electron laser FELIX.\cite{oomens2006gas} 
The light provided by FELIX consists of macropulses of about 5\,$\mu$s length at a repetition rate of 10\,Hz, which contain 0.3 to 5\,ps long micropulses with a micropulse spacing of 1\,ns.
The wavelength is continuously tunable over a range of 40 to 2000\,cm$^{-1}$. 
Here, typically wavelengths from 500 to 1850\,cm$^{-1}$ were scanned. 
When the IR light is resonant with an IR active vibrational mode in the molecule, this results in the absorbance of many photons, which causes dissociation of the ion (IRMPD). 
Monitoring of the fragmentation yield as a function of IR wavelength leads to the IR spectra.

\section{Acknowledgments}
The authors acknowledge continuous interest and support by Gerard Meijer (Radboud University Nijmegen).
We gratefully acknowledge the ``Stichting voor Fundamenteel Onderzoek der Materie'' (FOM) for providing the beam time on FELIX as well as support by members of the FELIX staff: Britta Redlich, Lex van der Meer, Rene van Buuren, Jos Oomens, Giel Berden, and Josipa Grzetic.
Franziska Schubert, Sucismita Chutia, and Mariana Rossi (FHI Berlin) are acknowledged for discussion and technical help.
C.B. is grateful to Hans-J\"org Hofmann (Universit\"{a}t Leipzig) for inspiring discussions.

The Supporting Material to this article contains further details of the simulation setup, experimental procedures, energy hierarchies for AAPA+\ce{Li+} geometries with and without the cation at different levels of theory, backbone torsion angles of low-energy conformers of AAPA and ADPA in isolation and in the presence of \ce{Li+} and \ce{Na+}, respectively, and Cartesian coordinates of the structures displayed in the article.
The material is available online at \url{http://www.fhi-berlin.mpg.de/~baldauf/publications.html}

\bibliography{Baldauf_2013_cations}

\end{document}